Linear laws of volume elasticity in the properties and structural phase transitions: physical process of the parameter effect (TPH) in magnet-structures.

*P.I. Polyakov*

*Mining Processes Physics Institute of National Academy of Sciences of Ukraine, 72 R.Luxemburg Str., Donetsk 83114, Ukraine*

*poljakov@mail.fti.ac.donetsk.ua*

*+38 (062) 311 52 85*

**Abstract**

The present paper deals with the analysis of experimental results taking into account mechanisms brought by the bulk elastic energy transformed by the thermodynamic parameters, temperature, magnetic fields, high hydrostatic pressure (T-H-P). An effect of the external parameters is considered through the separation of critical lines and points of the "cooling-heating" effect in the course of analysis of linear elastic evolution of properties in sign alternating and crossing effects of elastic stress energy as well as their value for structural reorganizations at reversible volume changes (structural phase transformations of types I and II).

An explanation of the direct and reverse hysteresis effect in the range of the structural phase transitions I and II is suggested and the secondary signs of changes of phase state properties in the wide range of structures are formulated. The regularities of formation of a structural phase transition at 0 K are stated with separating the position of the triple point and the change of the properties and phase states with the elements of superconducting and conducting properties.

We have marked out regularities of changes of $Eu_{0.55}Sr_{0.45}MnO_3$, $Sm_{1-x}Sr_xMnO_3$ properties and its isotopic analogue with oxygen substitution in the course of formation of the structural phase transition of type I at 0 K and evolution of the boundaries of phase states. We suggest new approaches for defining of critical lines and points in physical processes of the separation of the boundaries of phase states.

We have carried out the analysis of elastic linear changes of resistivity and the jump of properties under the effect of three parameters in polycrystal $La_{0.56}Ca_{0.24}MnO_3$. Using the examples of linear changes of resistivity, structure parameters, magnetization and magnetostriction of the structural phase transitions of types I and II in metals, semiconductors, dielectrics under T-P-H influence, the governing role of the laws of bulk elasticity is defined. The analysis described below underlines the dominating role of elastic internal stresses (EI stresses) with the energy that formed the dynamics of interactions determining the structure state, properties and critical phase transformations in magnet-containing media.

*Keywords:* Anisotropy; Pressure effect; Phase transition; Interaction mechanism

**1. Introduction.** At present, the number of papers studying of magnet-containing multi-component manganite systems increases due to high experimental potentialities. Some uncertainty presents in the form of anomalies, peculiarities, numerous effects with contradictory interpretation of mechanisms of the colossal magnetoresistance (CMR) realization. While studying magnetic–phase states, an attempt was made to relate the formation of the CMR state to changes of the bulk magnetostriction in single crystals of different manganites, as noted in reviews [1-2] and papers [3-6]. But it is not clear what phases and mechanisms are responsible for the changes. A number of anomalies, effects, features have been revealed for manganites modified by different chemical additives of varied percentage, e.g., giant negative magnetostriction, identity of modification of electronic properties influenced by both the temperature and the magnetic field H. Unfortunately, the role of phase transition (PT) has not been understood and grounded completely in the course of analysis and substantiation of the majority of anomalies and effects in properties where the main attention was paid to the determination of strong bonds between unfilled electron shells and to their reaction to the influence of thermodynamic parameters with magnetic properties taken into account. Moreover, an interpretation of the Curie temperature was indistinct to some extent. An interest to the correspondence between hydrates and manganites with respect to similarities and divergences of lattice and magnetic effects was shown in review [2] without interpretation of structural changes and the role of elasticity. The conception of two co-existing phases was used as the approach to the analysis of colossal magnetostriction phenomenon. The elastic energy was separated from the Hamiltonian and omitted, being replaced by the term of phonon degrees of freedom.

The results of investigations of peculiarities are demonstrative for magnetic and electric properties [5]. Here the colossal magnetoresistance in $Sm_{1-x}Sr_xMnO_3$ (x=0.45) was explained by high negative bulk magnetostriction, and changes in parameter (x) are considered as a correlation of magnetoresistance, magnetostriction, magnetization resulting in anomalies and singularities. The work [7] is of a special interest, where peculiarities of semiconducting regions with varying conduction current in $Eu_{0.7}Sr_{0.3}MnO_3$ were analyzed and an original explanation of a giant maximum on the isotherms within the low-temperature range was suggested. Besides, the differences were shown between magnetization at varying temperature of the sample with the field and without it, which are typical for the studied system from the viewpoint of the authors.

The role of the oxygen-isotopic substitution in the $Sm_{1-x}Sr_xMnO_3$ system was studied in [8], where peculiarities of the metal – dielectric transition and regularities of changes of the low-temperature phase of an unclear nature were substantiated.

In [9, 10, 16, 17], new approaches and explanation have been used to study T-P-H influence on properties and PT in single–crystal samples of low-temperature magnetodielectric, magnetic semiconductor and its analogues.

In the mentioned papers, the specified variety of critical lines and points underlined the causal basis of the mechanisms of ED stresses and their role in processes of a sign inversion, change of parameter priorities, and a sign inversion of thermodynamic parameter effect on the properties in different phase states. The direct correlation was supposed to exist between changes of the structure, magnetic and electronic properties; the dominating role of mechanisms of bulk EAD stresses was grounded.

Now the perfection of the used methods and the progress in model presentations resulted in the division of the experience into narrow areas where the general regularities of physical properties are hardly taken into account. We should pay attention to the problems that do not allow to separate real physical process in the course of the property evolution studies. Very often, the registration methods are not impartial and informative. For instance, the measurement of lattice structure changes under combined thermal (T) and magnetic (H) impact is impossible with using the standard X-ray methods. It was mentioned in the papers of P.L. Kapitsa [13]. That is why the identical PT in magnet-contained structures were separated on T-influenced structural phase transition and its magnetic analogue. A combined study of T and H effect was the study of magnetic resistivity in a wide spectrum of pure metals. Almost all the samples demonstrated linear regularity called the Kapitsa linear law. Basing on these investigations, the author supposed that in magnet-containing materials, magnetic fields introduce additional IE stresses into the lattice bonds that respond as linear changes of conductivity. At that time, the absence of experimental capabilities of the investigations under hydrostatic pressure did not give an opportunity to prove finally the mechanism of IE stresses influencing lattice changes under magnetic field unto 350 kOe within the temperature range of 10-250 K.

For instance, a similar linear regularity was presented for the resistivity under T-P-H parameters from 80 to 250 K in magnetic fields to 8 kOe and the hydrostatic pressure up to 18 kbar in $La_{0.56}Ca_{0.24}Mn_{1.2}O_3$ [9]. We should note that it was the influence of all the parameters (T-P-H) that resulted in a change of the properties in analogous linear regularities of elastic stresses which allowed us to find the correlation between the control parameters (6.8 K – 2.8 kOe – 0.1 GPa). The results of the measurements of the second type structural phase transitions and the linearity of magnetostriction (bulk striction) in $LaMnO_3$ single crystals within the temperature range of 70-200 K in the fields up to 300 kOe [6]. Here almost all the properties before and after the phase transition had the adequate regularity that allowed us to define the correspondence T-H (5.7 K ~- 2.5 kOe). The same linear regularities of magnetization were revealed in the studies of oriented single crystal magneto-dielectrics $CuCl_2 \cdot 2H_2O$ at temperatures between 1.5 K and 4.0 K in magnetic fields up to 20 kOe by Poulis [25]. Further investigations of resonance properties in the varied temperature conditions of 1.57-4.2 K and magnetic fields up to 10 kOe and hydrostatic pressure to 11 kbar [23] in the wide frequency spectrum had shown the linear dynamics of evolution of the structural phase transition of the first type. With all this going on, phase states are divided into the changes of magnetization and resonance properties with fixing the regularities of the jump of the properties and the peculiarities of the states. Basing on the frequency-field characteristics, the role of competing thermal stresses and additive magnetic ones is depicted in the formation of priorities of electron states of corresponding bonds in the structures that are responsible for both the conductivity and resonance absorption (the variation of resonance frequencies). The same structures were chosen to show the role of elastic stresses in the processes under hydrostatic pressure up to 2 kbar studied using X-ray method [11]. For $CuCl_2 \, 2H_2O$ at room temperatures, the linear change of structure parameter was found and the variations depending on the anisotropy of elasticity were discussed. The linear role of bulk elasticity was the most demonstrative in the course of X-ray measurements in $La_{0.875}Sr_{0.125}MnO_3$ [19] and polycrystals [26]. The estimation of linear changes of parameters in wide temperature range of 300-1500 K clearly demonstrated the role of thermoelastic stresses repeated in $LaMnO_3$ polycrystal structures and in $La_{0.7}Sr_{0.3}MnO_3$ at the temperatures of 300-1100 K before and after structural jumps.

Unfortunately, methods of investigations under the hydrostatic pressure were far from being perfect at that time. Our result shows only a combined T-P-H effect through mechanisms of ED stresses. It provides an objective consideration of physical processes, otherwise, new approaches cannot be elaborated. Far in the past, the lack of experiment could not give an impulse to the development of model-theoretical representations to substantiate real physical processes.

The object of the present paper is an analysis, drawing analogies and generalization of regularities of an evolution of structural phase transitions and properties in multi-component structures of magnet-containing systems of magnetic dielectrics, semiconductors and metals. It is also aimed at the presentation of unambiguous T-P-H impact through mechanisms of IE stresses and settling the question of the presence of anomalies and regularities in view of the laws of the bulk elasticity. We will also specify the peculiarity of a structural phase transition of the first and the second type at T=0 K and an opportunity to treat in a new fashion these critical phase states.

## 2. Results of the analysis and generalization of experimental investigations.

2.1    While considering the properties of solids as a set of incoming chemical elements that form the lattice structure,  dynamical lattice changes are revealed as the evolution under T-P-H influence at the formation of the structural phase transitions and phase properties. The conditions determining the locations of atoms at lattice sites are high-temperature physical and chemical processes of the structure formation. Numerous further studies of physical processes and models are connected with the methods of the realization and the search for the major driving force (energy) determining mechanisms of the changes, inner interactions, structure, phase transitions and properties where the processes of structure formation and elasticity are not often taken into account.

As shown by the specific dynamics of the changes in the course of the T-P-H effect, we can establish and separate the causing role of long-order anisotropic IE stresses that form structural phase transitions of displacement and rupture.

We consider the models of the nature of interactions binding atoms and ions at the sites of the lattice structure where the determining role is played by free and valency electrons. Those are peripheral electrons with the energy lower by two or more orders that bind a set of atoms and ions within a lattice in the course of structure formation. Being transformed by heat sink, the process thermal compression changes the ratio of all energetical states of electron interactions and forms both the conducting and magnetic properties.

Basing on the fact that the dynamics of changes is revealed in processes of elastic deforming influences on the relation of energies reflecting the structure state in the properties, phenomena, effects, we should accept that all energies with the comparable contributions to the processes of interaction have to take part there [18]. The average estimations of the interaction energies are listed below.

1. Exchange energy of electron interaction with the width of electron zone of 1-10 eV.

2. Energy of bond with the crystal structure (0.1-1 eV).

3. Spin-orbital bond energy estimated as $10^{-2}$-$10^{-1}$ eV.

4. Magnetic spin-spin bond of $10^{-4}$ eV.

5. Energy of electron –nuclear interactions of $10^{-5}$-$10^{-6}$ eV.

6. We should add the energy of elastic deformations (i.e. stresses deforming the lattice) here, that is estimated [23] as 1-200 GPA accounting for the coefficient of elastic compliance that corresponds to the energy of 0.1-20 eV.

The estimations allowed us to suppose that the energy of elastic stresses is more comparable with the energy of exchange interactions and is the major energy parameter determining the bonds and states of all interactions to be less important. As a result, the IE stresses influence the structure changes and determine the commonality of mechanisms of the effect of the parameters (T-P-H) on the evolution and formation of structural phase transitions, critical states and properties. That fact excludes the existence of an equilibrium state of the studied object. Thus, the influence of the energy of elastic stresses on the interaction energy modifies the energy of electron bond between atoms of the structure. We should keep in mind that the energy of dense bonds between the structure sites at definite conditions forms the magnetic properties, and the energy of free electrons is related with the state of conductivity.

We should take into account that the energy of elastic deforming stresses is seen in linear regularities of resistivity, magnetization and striction.

Such a position simplifies understanding of the main laws of interactions with respect to the analysis of the experimental results of concrete physical processes.

First, it is necessary to make qualitative definitions in the analysis of the principles forming the creation of a structural phase transition and properties of phase states. We should understand intuitively the relation of experimental results and evolution of a physical process from the viewpoint of phenomenology where the reaction of tensor elastic properties of structures is reflected that are formed by interaction energies and correspondence with thermodynamic parameters.

a)      the first qualitative definition is the transformation process where one kind of energy is brought in and another one functions. For instance, heating results in additional structural IE stresses through the mechanisms of thermoelasticity. The inverse process is cooling with the mechanisms of thermoelasticity reducing the level of elastic deforming stresses. Transformation of temperature into elastic deforming stresses changes all the energies by means of structure interaction and is the determining factor of structural phase transitions and phase states.

b)      The influence of hydrostatic pressure is transformed into the structural changes analogously to the cooling process (heat sink). At the rate of backpressure, the level of elastic deforming stresses is reduced due to barraelasticity and the structure is changed. The inverse result is achieved by the pressure reduction analogous to the heating of the structure. The transformed elastic deforming stresses are enhanced here. The level of changes brought up to the structure depends on both the loading limits and the coefficient of elastic ductility. The value of the coefficient varies from $10^{-2}$ down to $10^{-8}$ in different structures.

c)      Further we consider an effect of the magnetic field. It is necessary to underline that there is no magnetization ability of the structures predisposed to paramagnetism at high temperatures here. The temperature is a structure formation factor determining the energy of bonds and interactions. The succeeding cooling and the loss of heat energy are transformed in the structure (as shown by X-ray method [19]) into the linear elastic change of the parameter and the volume which witnesses for the mechanisms of elastic deforming stresses. Energies binding the atoms and the sites in the structure and forming the crystallographic succession are of both electrodynamical and elastic origin. Their joint action reflects the state of the structure in physical processes where the parameters influence the evolution of the volume and the form. That is why the bonds of electron density are formed responding to the sufficient amount of uncompensated magnetic moments that are lower by definition. This process is accompanied by the definite thermal conditions forming the structure by the rate of EDS up to the states and interactions where the effect of the magnetic field on the uncompensated moments is realized as magnetization. The succeeding rise of the level of lattice changes under magnetic field results in the bulk magnetostriction. We should note that the configuration of the electron orbits connecting structure sites is anisotropic by definition. General qualitative presentation of heat sink, temperature drop P and H influence is transformed into the long-order energy of ED stresses that results in redistribution of all lower energies of interaction on the background the structure lattice changes. Thus, the conditions transformed into IE stresses and affecting the structural phase transition at the corresponding T with the added magnetic IE stresses are seen as the evolution of dynamics of a phase transition with the fixed Curie temperature ($T_c$). We should add that the limited temperature area of formation of sufficient uncompensation and the conditions of the corresponding structural phase transition results in the jump of the specific heat capacity, structure parameter and symmetry, susceptibility, magnetostriction, magnetization and magnetoresistivity.

Of course, the list of the qualitative presentations is not full but it presents a principal and fundamental method of analysis of the physical processes, being based on the generalization of a large quantity of experimental results dealing with critical phenomena of structural phase transitions and the properties of phase states.

2.2    Numerous studies of copper chloride dihydrate $CuCl_2 \cdot 2H_2O$ [9,10,16] were taken as a basis for drawing analogies and correspondences in mechanisms and regularities of changes of structural phase transitions and properties of the mentioned multi-component magnet-containing systems. The interrelation of elastic mechanisms was established for the joint thermodynamic influence of P-T-H on multi-component magnet-containing structures.

Using the method of analogies and comparisons [10], we have brought to conformity the dependencies of magnetostriction in $LaMnO_3$ and magnetization in $CuCl_2 \cdot 2H_2O$. We have emphasized the causing role of thermoelasticity and magnetoelasticity in linear regularities of the properties (Figs. 1a, 2a). We should note that the same features illustrate one more important regularity of identical effect of T and H on both magnetostriction dependencies (Fig. 1a) before the phase transition in $LaMnO_3$ and magnetization (Fig. 2a) after the phase transition in $CuCl_2 \cdot 2H_2O$. We state that the separated areas have the same mechanisms of influence and belong to the conducting (metallic) phase of both the structures in the region where changes of the properties under T and H are antiphase. The next important result is established in magnetization behavior at the initial stage (Fig. 2a). We have singled out stable tendency of the change of properties with respect to the critical point $PP_x$ at similar T and H influence and the regularity of the motion of the critical line $T_p(H)$ of the temperature-field dependence was established that gave us an opportunity to demonstrate the dynamics of structural phase transition of the first type at T=0 K with the definition of the critical point $P(T_p=0 K, H_p=5.5 kOe)$.

These results form new approaches to the analysis of phase diagrams [10, 13], low-frequency properties both at 0 K and at selected temperatures in $CuCl_2 \cdot 2H_2O$. New methods of analysis using approximations of the corresponding curves based on the elastic linearity don't contradict to the logic of the studied reversible processes and allow to define the position of a structural phase transition of the second type at 0K both in $LaMnO_3$ $T_x(H_x \sim 200 kOe, T=0k)$ and in $CuCl_2 \cdot 2H_2O$ $T_x(H_x=5.5 kOe, T=0 K)$, Figs.1a and 2a.

As there is a problem of obtaining experimental data exactly at 0K, it is very important to establish general signs of the change of properties near this temperature (further we shall call them the secondary signs) that follow the phase transition and are formed under T-P-H influence. The presence of these signs in other systems and samples will define the generalizing role of the mechanisms of IE stresses in the structural changes and will allow us to single out these regularities in a large number of papers where they are treated by authors as effects, anomalies and peculiarities.

We should add that we pay attention to the relation of elastic magnetic properties and the orientation of the structure with respect to H. Thus, we can find the position of the critical points $P_x$ and $PP_x$ on Figs. 1a, 2a. Both magnet-containing and the structural changes connected with the composition and technology of preparation of the samples shift the location of the critical points $P_x$ and $PP_x$ (Fig. 1a,b) and critical lines in the form of temperature-field dependencies of the structural phase transition of type II of T(Hg) that is the boundary of separation of the metallic and semiconducting phases. This is the reason why the location of coincidence of $PP_x$ and $P_x$ (Fig. 2a,c) and the role of sign alteration in the position of the critical line of temperature-field dependence of the structural phase transition $T_p(H)$ show that the critical line separates both the conducting (metallic) phase and the phase where T and H influence the magnetization analogously. The result of thermoelastic compression as a regularity of antiphase thermal influence reflects both the peculiarities of the structure and the anisotropy of ED stresses. This fact plays the causing role in formation of visible changes of the properties in the form of the jump in the region of the structural phase transition at 0K. We should note that the discussed regularities found in one model structure are repeated in a number of multicomponent systems where the mechanisms of the influence are analogous and follow these conditions of formation of structural phase transitions and the evolution of the properties.

We should say that the consideration of the T and H effects through cooling- heating effect of EAD stresses with the account for the role of the selected critical lines $T_p(H)$ and points $P_x$, $PP_x$ gives a possibility to establish the mechanisms of the change of properties in direct and reverse hysteresis in the course of the analysis of experimental results.

When considering the direct hysteresis (Fig. 1a,c) in magnetostriction dependencies in the area of the phase transition in $LaMnO_3$, we paid attention to the value of thermomagnetic IE stresses in the influence of thermodynamic parameters T and H [9,10]. In these papers, the regularity of the influence of P and H through the cooling-heating effects and sign alteration is illustrated as well as the change of priority values of thermal and magnetic IE stresses in the changes of structural PT, magnetic and resistive properties. We have found the correspondence between the evolution of magnetization in phase states (diagrams) of low-temperature dielectric $CuCl_2 \cdot 2H_2O$ [16].

These results and the position of the selected critical lines and points in the studied properties of $LaMnO_3$ put a question about their role in the changes of hysteresis near the structural PT.

In $LaMnO_3$ magnetostriction (Fig. 1a), the role of the "cooling" effect of the magnetic field at the initial stage is connected with the regularities of $P_x$ critical point. We should note that the evolution of changes of the properties under H influence is antiphase to the temperature change. The nest regularity is the behavior the dependence after PT in a new phase state where the change of the properties follows the changes with respect to the critical point $PP_x$ (Fig. 1c).

In this region, the heating effect influences the changes at the reverse and the taking down the magnetic field. The jump of the properties results directly from the changes within the sample relatively to the location of $PP_x$ critical point where the increase of the temperature is followed by the return of the properties to the initial phase (Fig. 1a,c).

This physical process is the direct hysteresis in the area of structural PT and it is depicted with the account for the influence of EAD stresses with the fundamental role of the critical lines and points.

The next example of the change of magnetization in $CuCl_2 \cdot 2H_2O$ (Fig. 2a,c) allowed us to establish a regularity corresponding to the effect of reverse hysteresis and to define it accounting for the locations of the critical points $PP_x$ and $P_x$ and the changes of temperature – field dependence $T_p(h)$ of the structural PT. The result shows that here the critical line $T_p(H)$ separating the phases is attached to the position of $PP_x$ (Fig. 2b) and the change of the properties (Fig. 2a) at the initial stage of the magnetization with the single-value influence of T and H is the consequence of the cooling effect of the magnetic field and fixes T changes relatively to the critical point $PP_x$ (Fig. 2c).

The further evolution into another phase where the influence of T and H is antiphase fixes the jump of magnetization with the increase of H (Fig. 2a,b) in metallic phase and relatively to $P_x$ (Fig. 2c). Consequently, at the succeeding taking down of magnetic field H and at the rate of the elastic expansion, the heating effect is revealed on the magnetization relatively to $P_x$ with the transition to another phase with respect to $PP_x$. Thus, the properties are under single-valued influence of both T and H. This result is widely spread and follows to the established regularities.

This conclusion is an example of a definition of the well-studied effect of the direct and the reverse hysteresis that underlines the main reason of cooling-heating effect in the change of properties near the structural PT at rigid orientation and anchor to the position of the critical points $PP_x$ and $P_x$. Such treatment is a conclusive proof of the role of mechanisms of IE stresses in realization of the studied differences and can be an example of the presence of cooling-heating effects and the secondary signs proving the causing role of IE stresses in the regularities of the direct and reverse hysteresis.

As shown by the analysis of magnetization curves with respect to the critical points $PP_x$ and $P_x$ (Fig. 2a) before and after the PT, the temperature influence on the properties changes the sign. It means that at the initial stage, magnetization grows with the increasing T and H (all the changes are connected with $PP_x$). As was mentioned, such changes of the properties are attributed to the phase where magnetization changes in single phase with H and T that explains the jump of magnetization of the Meissner effect. That is a sign of a superconducting phase state. Analogous regularities are observed in $La_2CuO_4$ [27]. A varied set of experimental results is presented in the paper that render the studies of properties of stratified types of cuprates and rare earths in a wide spectrum of parameters that allowed us to include the mechanisms of EAD stresses into the consideration of formation of structural changes. We distinguish the role of elastic stresses and anisotropy of elasticity in the formation of structural phase transitions that are realized as the jumps of heat capacity, magnetization, elastic constants, density of concentrators of current carriers. The interactions induced by the energy of elasticity are seen through the correlation of electronic and magnetic properties that are followed by the separation of superconducting and metal phases in the course of the change of parameters.

The most unbiased results of studies of specific resistance of superconducting single crystal based on cuprates were presented in [27]. Within a wide temperature range, linear regularities of metallic phase of varied composition were established.

The variety of the enlisted works includes new understanding of the mechanisms in the causing role of volume elasticity. As a consequence, a structural phase transition is the regularity of elasticity influence on the formation of the area of critical states at the realization of the superconducting phase. In other words, the jump of the volume, specific density, elastic constants forms the structure of the optimal density which obeys the condition of optimal distances and bonds and the maximal density of untied free electrons responsible for ideal conductivity and superconductivity.

The next region of the curves starting from $P_x$ demonstrates antiphase influence of T and H that responds to the metallic phase state. The increase of the magnetic field enhances and the temperature reduces the magnetization. This is a characteristics of both magnetostriction of $LaMnO_3$ and magnetization in $CuCl_2 \cdot 2H_2O$. The changes of the properties in the selected regions belong to the metallic phase. As a result, $T_p(H)$ critical line (Fig. 2b) is the boundary of phase states where the temperature influence is sign-alternating. This result can be ascribed to the secondary signs.

Drawing the analogies of magnetostriction of $LaMnO_3$ and magnetization in $CuCl_2 \cdot 2H_2O$ and sign alternation in $T(H_g)$ (Fig. 1c) and $T_p(H)$ (Fig. 2b), we can state:

a)      In the first case, the temperature-field dependence $T(H_g)$ separates metallic and semiconducting phases;

b)      In the second case, the realization of the structural PT at $T_{cmp}=0$ K and field-temperature dependence $T_p(H)$ is the boundary of superconducting and conducting phases ;

c)      Critical line $T_p(H)$ (Fig. 2b) generalizes the changes of the structural PT after the jump of the properties and the effect of thermal elastic deforming compression (Fig. 2a,b) in the course of the cooling effect of the magnetic field is finished at the stated temperature $T_p=9.2$ K. The position of the critical point $P(T_p, H_p)$ fixes the boundary position of the SPT defined as the intersection of the approximated critical lines $T_p(H)$ and $T_p(H,P)$.

Separation of the secondary signs in model single crystals at the reversible physical processes in the varied magnetic field and temperature under pressure, the establishment of linear elastic dependencies of the positions of critical points $PP_x$, $P_x$ in the change of the properties allowed us to state that the influence of thermodynamic parameters transformed into EAD stresses is the major factor of formation of PT and properties. Critical temperatures of the SPT transformed into the level of ED stresses and the anisotropy of the elasticity shift the structural positions of atoms (sites) that changes the symmetry of the crystal (low-symmetry phase) on the background of the volume change and resulting increase of electron density. The critical phenomena of dynamically instable lattice result in the changes of the crystallographic structure and the following jumps of the volume, density, magnetization, heat capacity, resonance properties [24].

The established regularities and correspondences in magnetization of $CuCl_2 \cdot 2H_2O$ (Fig. 2a) and magnetostriction in $LaMnO_3$ (Fig. 1a) as well as sign alternation in changes of the locations of critical lines $T(Hg)$ of STP of type II with the defining negative sign (Fig. 1c) and $T_p(H)$ of STP of type I with the positive sign (Fig. 2c) are characteristics of phase states. These results allowed us to make the following generalizations:

a)    T(Hg) (-) regularity of field-temperature dependence of STP of the second type at $T_{uc}=T_{pp}$ and corresponding Tc demarcates metallic and semiconducting phases. Magnetic properties demonstrate antiphase influence of T and H and the effect of direct hysteresis.

b)    $T_p$(H) (+) field-temperature dependence of STP of the first type at $T_{ct}=T_{pp}=0K$ separates the metallic phase and the phase with elements of superconductivity where with single-value T and H influence , the effect of reverse (anomalous) hysteresis is realized.

While analyzing the established regularities and secondary signs, we used the results of a number of papers [5,6,7,8,15] where these anomalies were presented. We should note that experimental results of the chosen compositions of manganites are regarded to the study of multicomponent magnet-containing single- and polycrystal samples.

2.3    All the listed results with the important secondary signs give an opportunity to single out the role of T –H influence through the mechanisms of EAD stresses and confirm the significance of the critical lines and points selected with respect to the bulk elasticity mechanisms.

For the analysis, the most demonstrative are experimental results for manganites of combined $Sm_{1-x}Sr_xMnO_3$ composition [5]. The authors consider dependencies of changes of properties for samples with x=0.45. Magnetization isotherms have a singularity in the form of a jump of properties at the initial section of the dependence for fixed temperatures. With H increasing, a relative linearity at the region of saturation tends to maintain the magnetization, while the process of temperature growth changes the properties towards decreasing. The authors state that the maxima of conductivity on $\rho$(T) dependencies change by several orders of magnitude at low temperatures for all the investigated compositions, and magnetoresistance of samarium manganites of the taken composition (as stressed by the authors) reaches colossal values in relatively low magnetic fields. For instance, the magnetoresistance makes 44% for a composition with x=0.45 and H=0.84 kOe. The attention is attracted to a stable regularity of changes at isotherms with a jump of properties at the initial section of the dependence for all H fixed. The conformities retain in the whole temperature range up to $125^oK$. With changing T, the magnetization decreases in the region of the jump, i.e. with T increase in high magnetic fields, the magnetization diminishes linearly.

This result has made it possible to characterize and to substantiate the presence of general secondary indications in changes of properties of that structure:

a)      the inverse hysteresis observed in the region of phase transition allows us to treat the changes as a structural phase transition with the typical field – temperature dependence $T_P(H)$ with changes occurring at $PP_x$ critical point (Fig. 3);

b)      stable changes of magnetization isotherms after the jump of properties corresponding to PT, where the magnetic field enhances the properties whereas the temperature growth results in a significant decrease of the properties. It should be noted that this region of the phase state corresponds to the metallic conductivity;

c)      regularities of the magnetization increase under T,H influence observed on Fig. 3c show that at the initial section prior to the structural phase transition, the changes of properties under T and H effect occur in one phase (in view of position of $PP_x$ critical point). After the PT, the magnetization changes in antiphase to relative changes of T and H [5]. This result repeats changes in the magnetization for $CuCl_2 \cdot 2H_2O$ (Fig. 2a,b).

For further substantiation of regularities of IE stresses in the formation and changes of structural phase transition and properties, we analyze magnetic, elastic, magnetoelastic, resistive and magnetoresistive properties of polycrystalline $Eu_{0.55}Sr_{0.45}MnO_3$ [7]. The paper presents a large bulk of experimental results. The authors show that the studied ceramics is a polycrystalline single-phase perovskite of orthorhombic structure and, what is very important, they pay attention to the identity of the results to the properties of ceramic samples of other compositions: $Sm_{1-x}Sr_xMnO_3$

(x=0.4, 0.45), EuMnO$_3$. Hysteretic changes in magnetization jump within the temperature range of 1.4$^o$K-150$^o$K are explicitly seen on isotherms. The run of inverse (anomalous) hysteresis analogous to the behavior of magnetization for CuCl$_2$□2H$_2$O (Fig. 2c). The authors suppose that the structure remains dielectric in a wide range of low temperatures according to the resistivity properties. This important notice enables us to assume that the structural phase transition at 0 K can be a triple-point regularity and magnetic field H displays a jump in magnetoresistivity through mechanisms of magneto- IE stresses due to superconducting phase realization.

It should be noted that the resistivity should be much beyond the limits of changes in resistance with T decreasing, as follows from the logic of the experiment [7]. Most probably, the sensitivity of measurements was not high enough to register the peak of $\rho$ change at a level of $10^{10}$-$10^{12}$ at temperatures tending to 0 K. And it is shown in fact that the temperature dependence of thermoelastic expansion is practically linear with the low-temperature measurement errors being taken into account.

The above result makes it possible to note one more regularity shown on T and H dependencies of magnetization [7]. Here we see a jump-like increase of the magnetization at the initial section of isotherms with H fixed as well as magnetization decrease with T rising at the final part of the dependence. These changes are similar to those on the isotherms of Fig. 2a. They are the regularities of unambiguous change of the magnetization as a function of T and H at the initial section of the corresponding superconducting phase with $T_P(H)$ critical line separating phase states and corresponding to PP$_x$ critical point (Fig. 2b). The subsequent evolution results from the antiphase impacts of T and H in the conducting (metallic) phase, i.e. there is a sign-alternating temperature effect after the phase transition. In CuCl$_2$·2H$_2$O, the behavior of properties is similar (Fig. 2a). The changes observed on magnetization isotherms make it possible to consider this result as a secondary indication for this composition.

The analytical results emphasize peculiarities of formation of structural phase transitions and make it possible to construct and to single out $T_P(H_g)$ field – temperature dependence basing on changes of parameters of inverse hysteresis, Fig. 4. It is a regularity attached to the position of $PP_x$ critical point. The critical line of $T_P(H_g)$ separates two phase states (superconducting and conducting (metallic) phases). It also displays the regularities of structural phase transition at $T_{st}=0^oK$ and it is a dependence of the dynamics of changes influenced by the position of the averaged magnetic field $H_g$ determined in this paper.

It follows that the role of field and temperature effect through mechanisms of EAD stresses is a causal basis of volume change and the formation of structural phase transition and properties, as a consequence. Moreover, the position and sign alternation of critical lines and points determine the regularities of the impact in the form of changes of properties in both phases separated by the structural phase transition. Such an explanation implies that the role of elastically deforming stresses is the basic one as the binding energy directly corresponds to and interacts with elastically and anisotropically deforming mechanisms of volume change in this case. They are also the primary cause of PT changes and properties at the expense of mechanisms of thermoelastic and magnetoelastic effects of T and H.

An extensive experimental result [5] of studies of phase states and peculiarities of electrical properties in manganites of $Sm_{1-x}Sr_xMnO_3$ system was also treated in [8] in the course of investigations of a similar sample. The authors believe that the peculiar isotopic oxygen substitution helps in the identification of the metal - dielectric phase states. They pay attention to a composition with x between 0.4 and 0.5 where a jump of electrical resistance and sudden changes of the coefficient of volumetric expansion and magnetostriction were found out. High sensitivity of all physical characteristics to negligible composition changes, the lack of information for the explanation of properties of the above manganites have made us to propose the analysis of our own in view of the role of T-H-P influence through IE-stress mechanisms. We take the data for $Sm_{1-x}Sr_xMnO_3$, x=0.5. For the sample with x=0.5 and isotropic substitution, there is a phase transition separating metallic and dielectric phases within the low-temperature range. And a relatively weak influence of external magnetic field on magnetostrictive dependence [7] via mechanisms of magnetoelastic compression brings the system to a new phase state and, what is important, the corresponding maxima of the electrical resistance have much decreased and shifted to the region of higher temperatures under the influence of H. The temperature hysteresis decreases as well. As the result of resistivity change is incomplete, we show changes of $\rho$ for T=0 K (the dot-and-dash line) to be of the order of $10^{11}$-$10^{12}$ by approximation. This behavior of the dependence makes us to ascertain a regularity, that changes in thermoresistivity under magnetic-field effect (H=1T, 2T, 4T) fix the position of temperature maximum at $T_{PP}$ point. The specified effect corresponds to $T_{PP}$=0 K on the dependencies (Fig.5a). So, we can state that the structural phase transition is realized at 0 K in this sample.

It is also noted that the maximum MR is shifted in magnitude attaining colossal values for the mentioned samples. So, we can construct the field – temperature dependence of changes in PT showing the behavior of $T_P(H)$, Fig.5b. It is the critical line separating superconducting and conducting phases with the position directly related to $PP_x$ critical point responding to a real physical process. The authors stress that both the peculiarities of crystalline structure and the obtained results are similar to properties of single-crystalline samples of $Sm_{0.55}Sr_{0.45}MnO_3$ composition [14] having the rhombic space symmetry as well. The authors note that there are no structural phase transitions in the temperature range of 1.4-300 K but this statement is not entirely correct. It follows from our conceptions and secondary indications of $T_{PP}$-const=0 K rule that the pointed critical temperature fixes the structural phase transition and T=0 K=$T_{PP}$ responds to the position and properties of a triple point. The real character of the structural phase transition is $T_P(H)$ field-temperature dependence and changes of magnetic and electronic properties are consequences of peculiarities in different phase states revealed by the analysis.

All the listed results with the important secondary signs give an opportunity to single out the role of T –H influence through the mechanisms of IE stresses and confirm the significance of the critical lines and points selected with respect to the bulk elasticity mechanisms.

The same results are obtained after the manifold study of anomalies of magnetic and magnetoelastic properties of a single crystal $Nd_{0.55}Ca_{0.45}MnO_3$ in intense magnetic fields [15]. Reverse magnetic hysteresis was fixed as well as reduction of magnetization with the increasing temperature. These facts correspond to the secondary signs of the property changes of the phase state separated by the STP at 0K.

With important factors of secondary indication taken into account, the given results make it possible, first, to show the role of T-H impact through mechanisms of IE stresses and, second, to confirm convincingly the importance of the discussed critical lines and points and the role of bulk elasticity.

2.4    Consecutive analysis of the effect of IE stresses on the formation of structural phase transition and magnetic properties of the mentioned magnet-containing media raised a question, how the properties of conductivity behave at any changes of structure influenced by IE stresses.

The demonstrative are results of [9,16] considering mechanisms involved in formation of the dynamics of resistive properties in magnetic semiconductors. The authors studied the behavior of resistivity in a bulk ceramic sample $La_{0.56}Ca_{0.24}Mn_{1.2}O_3$ as a function of three thermodynamic parameters T-P-H (Fig.6a). The originality of this paper is that three thermodynamical parameters T-P-H are studied simultaneously. The merit of such a method of the study of resistivity is the most apparent for the mechanisms of realization of physical processes. The changes taking place in non-magnetic chamber of high pressure at hydrostatic conditions are related to the process of structure transformations, and optimal ration of the volume of the sample and the liquid is 1:100 to within P~0.1%.  The selected magnetic structure and its magnetic component are sensitive to the parameter influence with respect to the structural changes of interatomic bonds. Fig.6a presents the conductivity at the varied thermodynamic parameters. The decrease of the temperature (heat sink) results in the change of resistivity in semiconducting and metallic phase separated by the PT/ The succeeding influence of ED in the form of P (pressure effect) and magnetic field H with accounting for the critical point $P_x^1$ in the whole temperature range reduces the resistivity and changes the location of Tmc(H) and Tmc(P). Thus, magnetoelastic and baroelastic parameters influence interatomic bonds of the structure through elastic stresses that is revealed as critical phenomena in the form of resistivity jump. The dynamics of their changes is affected by the PT and the differences of anisotropies of elasticity and magnetoelasticity . The character of the process depends on the elastic ductility and the level of the transformed stresses.

The established physical processes are repeated in other results of the investigations that is characteristics of the structure changes at the rate of the energies of elastic volume stresses. That is external impact through the elasticity that transforms the interaction of inner energies making changes of the structure and forming structural PT and phase states. In addition, we shall establish the role of cooling-heating effects of P and H. The ratio of the parameters T-P-H is 6.8 K- 0.1 GPa – 2.8 kOe. These values for the given sample are obtained basing on the linear regularities of the experimental curves. Let us consider the effects using the triangle BCd and $AC^1d^1$ (Fig.6a). We trace the position of B point at the fixed P that is 1.81 Pa. The resistivity is reduced analogously to D with the same properties that is similar to the reduction from C with the corresponding heat sink about 120 K. The inverse effect is the taking down of P to zero corresponds to the increase of resistivity at the rising temperature. All these changes result in structural transformations. The analogous result gives the influence of the magnetic field H~8 kOe with the shift of C point to B` similar to the temperature reduction of 20 K. The inverse effect with the absent magnetic field H=0 is followed by the increase of the resistivity analogously to the increased T.

Taking into account high sensitivity of electronic properties to insignificant structure changes, we confirm the supposition concerning adequate influence of thermodynamic parameters on the structural changes through the laws of volume stresses. Thus, it is very important to iclude into the model both structural changes and levels of IE stresses, i.e. the elastic energy transformed by the parameters in physical processes.

We should add that the processes of heat sink (cooling) affecting the volume of elastic stresses form the STP accounting for the anisotropy of elasticity. The sample changes the form and the symmetry in magnetic field and this fact results if the change of magnetization at the rate of magnetic energy. The energy of deformation is joined at the level of the atom shift in the lattice. We should note that the method of separation of the critical point $Px^1$ depends on the position of P and H parameters that affect the properties in one phase. The same behavior is observed on magnetostriction curves (Fig.1a). The influence of the magnetic field H is strictly limited by the temperature of 315 K and is revealed further in the change of resistivity with the reduction of the temperature. At this temperature, interatomic bonds of the structure increase the electron density of the bond of the structure forming the level of magnetic non-compensation. That is the state of the noncompenstation reacts to the external magnetic field with the structure derangement resulting in the change of resistivity.

As formulated, there is an important regularity in such processes, i.e. conductivity redistribution because of the relaxation of inner stresses during volume reduction. As the thermoelastic expansion is associated with considerable inner stresses, the influence of elastic and magnetoelastic compression induced by the backpressure diminishes the inner stress. The process clarifies that any volume change due to thermoelastic, baroelastic, magnetoelastic compression relates directly to the inner-stress reduction in the system, while a volume decrease is directly related to the enhancement of the influence of energy of atom-electron coupling, by definition.

On the dependence (Fig. 6a), the denoted important factor of a nearly double jump in the region of PT, where the phase states are separated, implies that changes have occured already in the semiconducting phase. Such changes of properties are based on the presence of structural phase transition in a reversible physical process. In this case, the regularity is the equality of temperature maxima of thermobaroresistive, thermomagnetoresistive and thermobaromagnetoresistive effects $T_{PP}=T_{mc}$ that corresponds to the temperature of the structural phase transition. It can be consequently stated that the regularity $T_{PP}$-const is the law of conformity conservation for elasticity and magnetoelasticity anisotropies, which was first proved in [10] (Fig.6b).

Next, the shown critical lines of $T_{ms}(P)$ and $T_{ms}(H)$ fix regularities of the phase state separation and dynamics of their changes under the influence of hydrostatic pressure and magnetoelastic compression. It is obvious that the field-temperature and thermobaric dependencies reflect mechanisms of EAD stresses and the important role of thermo- and magnetoelasticity anisotropy involving into the processes. It is stressed that under P and H effect, the temperature area of the fixed changes of resistivity is extended in semiconducting phase (Fig.6a) and the same regularity is observed on Fig.5.

Typical changes of properties of semiconducting phase allow us to single out $P_x$ point of intersection of the dependencies by means of approximation. The critical state in the form of PT defines the relation between T-P-H induced change of properties and the real physical process of elastic-compression mechanisms. It means that any changes in volume towards reduction, e.g. a decrease of temperature parameter, affect the regularities of changes of properties in a way similar to hydrostatic pressure P effect and magnetic field H effect through magnetoelastic compression. These mechanisms directly relate to volume changes in volume and correspond one to another. As a consequence, it can be stated that position of $P_x{'}$ critical point functions as a critical parameter of formation of interacting binding energy and energy of IE stresses in the system.

It follows that the physical process of volume decrease at the expense of the revealed mechanisms of T-P-H influence modifies the binding energy of interaction of the atomic-electron system, and the structural phase transition separates a symmetry volume in one phase from a symmetry volume in another new phase. The extent of changes depends on the level of thermo - IE stresses.

Methods applied for revealing regularities in the formation and changes of structural phase transitions for determining phase states enable us to show the role of magnetoelastic mechanisms in the colossal magnetoresistance effect by means of the results of resistivity changes:

a)      to begin with, we pay attention to the behavior of dynamics of dependencies of resistivity change under T-P-H effect (Fig. 6a). With changing T and P, we have a region on $T_{mc}(P)$ dependence marked by $T_A$ temperature. Following the change of pressure with respect to $P_x'$ critical point, it is stated that with P increase, including the "cooling" effect, we observe resistivity change for fixed $T_A$ and this corresponds to the semiconductor-metal phase transition. The process is fixed by a sudden (not so high jump) change of properties n $T_{ms}(P)$ critical line. Similar regularity of changes in resistivity properties is found at $T_A$ and in the "cooling" effect of magnetoelastic compression by magnetic field H. A sudden (with the value of the jump depending on the structure change) resistivity change will be observed during semiconductor-metal transition through the critical line of $T_{mc}(H)$, Fig.6a;

b)      the following result is the most demonstrative and the analogy can be drawn by regularities of changes of properties according to investigations of electrical resistance for $Sm_{1-x}Sr_xMnO_3$ [5]. With $T_A'$ fixed, the "cooling" effect results first in dielectric-superconducting transition and there occur more essential changes in this range with further H increasing, i.e. resistivity reduction under the influence of mechanisms of magnetoelastic stresses and transformation into metallic phase takes place. The process of the change of properties becomes apparent by a noticeable resistivity jump in the region of the phase separation where $T_P(H)$ critical line has an intersection.

That are these regularities which explain the behavior of resistivity isotherms P(H) in a wide range of magnetic fields. It is illustrated on Fig. 5a by dependencies for fixed $T_1=20^oK$, $T_2=60^oK$ in $Eu_{0.55}Sr_{0.45}MnO_3$. At the initial section for $T_1=20^oK$, the change of properties with H increase corresponds to regularity of the "cooling" effect, when it is obvious that resistivity $\rho$ is invariably high, so it is the dielectric phase. We note that there is a jump of resistivity of the field of 20 kOe [7] which conforms the field-temperature dependence of the structural phase transition toward the phase with a considerable reduction of resistivity at the field raising. During the inverse process of H decrease, is presented in resistivity properties the "heating" effect as a transition from the phase with single-value influence of T and H to the conducting (metallic) phase followed by changes in the region of dielectric phase with H→0. Thus, it can be noted that the mechanisms of magneto-EAD stresses relate directly to conductance and the changes of the phase states in the critical region are the primary cause of the colossal magnetoresistivity effect currently studied.

An analysis of changes of properties represented on Fig.6a attracts out attention to dynamics of resistivity change in semiconducting phase and to the position of critical point $P_x'$ (Fig. 6a) and dielectric phase on isotherms of electrical resistance of [14]. This result clearly shows the scheme of changes of properties in the course of volume reduction as a result of T decrease and P increase. Similar result is observed in the case of H effect. In other words, the influence of these parameters on structural changes occurs involving the elasticity. This influence is unambiguous and conformal. The correspondence allows us to take $P_x'$ critical point into consideration and to represent P and H effects by mechanisms of baroelastic and magnetoelastic compression backpressure. It should be noted that anisotropy of thermo- and magnetoelasticity plays the principal role in the formation of structural phase transition, and the anisotropy of IE -stresses creates conditions for volume decrease in a reversible physical process. It is also an important factor for concentration of stresses preventing the mobility of conduction electrons in both semiconducting and dielectric phase till the boundaries where the structural phase transition develops accompanied by jumps of heat capacity and properties.

A consecutive determination of the role of IE stresses (compression) reveals regularities for the discontinuity of structure in the range of PT in a reversible physical process [20]. There is also a jump of properties (heat capacity) and temperature increase giving negligible changes of properties due to high stresses developed in the system, as shown by our analysis. This is also indicated by the observed accordance of temperature-to-pressure correlations and by a comparatively low binding energy. At low temperatures, such changes occur in the form of a large jump of properties due to low energies of IE stresses but with enhanced binding energy.

2.5     The performed analysis makes it possible to generalize and systemise the results of ascertaining of secondary indications in regularities of magnetic, elastic and magnetoelastic properties in $Sm_{1-x}Sr_xMnO_3$ compounds [5], they are:

    a)        inverse hysteresis is fixed, the jump in properties and the tendency of magnetization decrease with temperature growth are observed;

    b)        unambiguous growth of magnetization properties under T and H effect (the initial section of the dependence) is present as well as sign-alternating change of properties under the influence of H and T after the jump in properties. Temperature parameter affects the properties of different phase states in antiphase;

    c)        the regularity of changes in the field-temperature dependence $T_P(H)$ is revealed by averaging the position of jump of properties with the typical role of the initial point relating to $PP_x$ under the temperature change in angular coordinates (Fig.3);

Similar results with the revealed regularities corresponding to secondary indications have been observed in studies of peculiarities of magnetic, elastic, magnetoelastic properties in $Eu_{0.55}Sr_{0.45}MnO_3$ compounds [7]. Here we note dependencies obtained at T=1.4 K, that are close to those for T=0 K:

    a)        the parameter was changing linearly within the whole interval of measurements due to of thermoelastic compression. The method of approximations makes us to show the maximum of resistivity at the level of $10^{11}$-$10^{12}$ in dielectric phase;

    b)        considerable and obvious changes of the inverse hysteresis were present on the magnetization isotherms including T=1.4 K, as well as the typical reduction of properties with T increasing;

c)      unambiguous growth of magnetization with T and H increase was observed at the initial section of the dependence followed by the antiphase effect of T and H at the final stage;

d)      the analysis of the dynamics of changes on magnetization isotherms (Fig. 4) makes it possible to mark and represent the regularity of the field-temperature dependence as $T_P(H_g)$ , i.e. as the critical line of the structural phase transition with $T_{st}=0$ K, separating superconducting and conducting (metallic) phases with the typical position of $PP_x$ critical point and respective change of the temperature.

It should be added that the observed character of evolution of properties and the determined critical temperature $T_{st}=0$ K meets the requirements set forth for the formation of superconducting and conducting phase. It is also a regularity satisfying the conditions of the triple point.

Hence, it can be concluded that one more critical point P with $H_P$ and $T_P$ coordinates can be taken into account for the samples under investigation. This point was shown first in [10] and determined by approximation and intersection of changes in the field-temperature $T_P(H)$ and field-baro-temperature dependencies $T_P(H,P)$ presented on Fig.2b. To determine such a point in studied structures, a cycle of investigations of properties and PT influenced by thermodynamic parameters T-H and by an important parameter (the hydrostatic pressure P) should be performed.

By extending potentialities of the analysis on revealing secondary indications, the same regularities can be found in the so-called anomalies of magnetic and magnetoelastic properties in $Nd_{1-x}Ca_xMnO_3$ single crystals in strong magnetic fields [15]. The represented dependencies of magnetization for composition with x=0.45 demonstrate inverse hysteresis of change of properties and make it possible to show definitely that changes of the field-temperature dependence of PT are in correspondence with respect to $PP_x$ critical point. It should be noted that magnetization decreases with temperature rise, and $T_P(H_g)$ field-temperature dependence is a consequence of temperature of the structural PT detected at $0^oK$. The results to be analyzed might be more reliable provided by investigations accounting for selected directions. The same conclusion follows from the analysis of properties and anomalies of changes of phase transitions in $La_{1-x}Sr_xMnO_3$ single crystals [6].

The presented methods of analysis, the established regularities of elastic stresses in physical processes of formation of phase transitions and properties in multi-component magnet-containing structures are separated among fair quantity of results and facts. The main goal of the present paper is the demonstration of universality of the chosen methods of analysis. This aim demanded the intuitive formulation of principles of correct account for macro- and mezoscopic properties.

The establishment of causing role of elastic stress energies and natural fact of their separation at formation of critical lines and points [10] attracted our attention to [1,8,15,21]. There a wide spectrum of studies of multi-component magnet-containing structures is presented in the form of varied parameter dependencies of phase diagrams of $Re_{1/12}Al_{1/2}MnO_3$ [1,21] compounds, $Pr_{1-x}Ca_xMnO_3$ single crystals and $Pv_{1-x}Sv_{1/2}MnO_3$ [8] and$Nd_{1-x}Ca_xMnO_3$ [15], too. Presented in [9,10] for $LaMnO_3$ and $CuCl_2 \cdot 2H_2O$ properties, the selected methods of analysis allow us to establish analogous regularities also in these papers. We should note that the singled-out T(Hg) dependence (figs. 8b, 9b) is the average result of the handling of changes of magnetization and magnetostriction. The identity of these properties shows their complete dependence on the mechanisms of magnetoelasticity.

The exhibition of $Pr_{1/2}Sr_{1/2}MnO_3$ phase diagram [8] (Fig.7a) as T(Hg) dependence with separated $P_x$ critical point and $T_x(H_x \ T=0)$ shows common regularities with $LaMnO_3$ (Fig. 1a,b). Only the positions of the singled-out critical points are changed. Similar dependencies can be established for the whole manifold of phase diagrams of the varied compositions of $Re_{1/2}Al_{1/2}MnO_3$ (Fig.7b). The character of the singled-out dependencies is analogous (Fig. 7a), except the position of the critical lines and points.

The succeeding result is presented in the form of a phase diagram of magnetization and magnetostriction of $Nd_{1-x}Ca_xMnO_3$ single crystal [15], Fig.8. The cited results have the form of temperature-field dependence $T(H_g)$ (Fig. 8) separating superconducting phase and metal analogous to phase transition in $CuCl_2 \cdot 2H_2O$, Fig. 2b.

The established position of the critical points $PP_x$ allowed us to separate the same regularities in phase diagrams of single crystal $Pr_{1-x}Ca_xMnO_3$ too, where small concentrations of the composition x=0.3; 0.35; 0.4 can be treated as a phase transition analogous to the corresponding structural phase transition in $CuCl_2 \cdot 2H_2O$. The succeeding changes of the composition x=0.45; 0.5 result in sign alteration of the position of the critical line [10] corresponding to the phase transition analogous to single crystal $LaMnO_3$ behavior (Fig. 1b, 7a) with the singled critical point $P_x$.

We should note that the magnetization of the superconducting phase changes under the influence of T and H in single-phase state.

The derived regularities are justified by the methods of the carried analysis where the causing role belongs to the energy of elastic stresses. If the volume elasticity is not taken into account in model presentations, it means the abandonment of the analysis of the role of elastic energy in formation and evolution of structural phase transitions and properties.

2.6 One of the direct methods of registration of structural changes is X-ray method. The registration of the heat sink process can be made by the change of the lattice parameter due to atomic bonds in the system of single crystal $La_{1-x}Sr_xMnO_3$ with the temperature increase [19]. The evolution of the lattice parameter for all the studied compositions is linear (Fig.9). The linear regularity of the lattice parameter as a function of temperature with an allowance for error  is observed in the whole range of the investigation from 373 K up to 1473 K in single crystal $La_{0.875}Sr_{0.125}MnO_3$ at the rate of thermoelasticity. The same paper presents the results of variation of the ratio La and Sr. Linear regularities stay stable within the whole temperature range. Here the jump-like change of the parameters is presented too, being determined by the presence of SPT in the range of 300-350 K. The transition from rhombohedral symmetry at 88 K to orthorhombic responses at 600 K is fixed. Thus, the cause of the changes of the lattice parameter is the bulk elasticity within the whole temperature region. It is transformed by the heat sink and gives origin to the mechanisms forming the condition of a PT ant the properties of phase states where magnetic properties are started from T~350 K.

Analogous X-ray measurements of polycrystal $LaMnO_3$ and $La_{0.7}Sr_{0.3}MnO_3$ have also shown the singled out linear elastic dependencies of the parameters in [26]. The presented linear registration stresses the significant role of DS in the bulk elasticity and the processes of heat sink (cooling). Here we should note that the magnetic properties are formed below 350 K in these systems.

To estimate the elastic parameters and regularities of a single crystal, the influence of the uniform hydrostatic pressure on the parameters of major crystallographic axes was studied. The measurements were done at room temperature in monocrystal $CuCl_2 \cdot 2H_2O$. The experiment was carried out using X-ray diffractometer at 0-2 kbar. The evaluation of compressibility in Co-radiation yielded the interfacial parameters in a linear form (Fig.10). The linear changes of the parameters and the volume compressibility are obvious proofs of the role of elasticity in structural changes.

As an example, we should draw attention to the works of Kapitsa [13] dealing with the bulk elasticity and the conductivity in metals. As shown by analysis of about 36 metals, the characteristic linear changes of the conductivity are stable within a wide range of field H. The typical examples are presented on the dependence of specific resistance of copper (Fig.11) at fixed $T_1$=86 K (1), $T_2$=63 K (2), $T_3$=20 K (3) with H varied from 0 to 300 kOe. The structure of Cu type possesses the magnetostriction properties in the considered temperature range in magnetic field and reacts on the influence of T and H parameters by the deforming violation of lattices. In the first place, it results in linear change of conductivity (Fig.11).

While considering the selected dependencies through the cooling-heating effects of the magnetic field, we can draw analogies of dynamics of A, B, C points, linear regularities of elasticity and the position of $P_x$ critical point with magnetostriction properties of $LaMnO_3$ (Fig.1a) in metallic phase state. The mentioned results define the causing role of linear elasticity in magnet-containing metals, too.

Taking into account the proven mechanisms of IE in realization of physical processes using real experimental results (significant jump and linearity of magnetization at T=1.53 K in $CuCl_2 \cdot 2H_2O$ (Fig. 2a) and inverse hysteresis and linearity at $T=1.4^oK$ in $Eu_{0.55}Sm_{0.45}MnO_3$ (Fig. 4)), we can believe theoretical models based on ideal postulates and containing significant restrictions to need additional detailed elaboration [18]. The existence of considerable binding energies verified by quantum-mechanical definitions side-by-side with EAD stresses forming a structural phase transition (a triple point) exactly at 0 K is the driving force of a real physical process. The evidence of the existence of a structural phase transition at 0 K is the availability of a number of secondary indications enlisted in a number of papers in the form of jumps of heat capacity, magnetization, resonance properties, striction etc. that emphasizes the real importance of the mechanisms of EAD stresses.

As a consequence, the formulated regularities contradict to the statement about the absence or insignificant value of IE stresses [18] at the formation of structural PT at 0 K.

While extending the areas of use of methodology of the analysis to different systems, we can show analogies and correspondences of the mechanisms of realization of the structural phase transition of type I in $CuCl_2$ $2H_2O$ and $MnF_2$ [10, 20, 23, 28]. In both systems, the investigations were carried out with accounting for the relative orientation of the structure and the magnetic field. Both magnetization (Fig.12c) and $T_{ct}(H)$ dependence (Fig.12b), the frequency-field dependence (Fig. 12d) and phase diagram (Fig. 12a) repeat the regularities of the same dependencies in $CuCl_2 \cdot 2H_2O$ (Fig. 2a, b, c, d) where the role of T, H, P influence is transformed into the processes of elastic deforming stresses. The difference of the structures of the mentioned single crystal realized in the coefficient of the elastic ductility with the value of $10^{-3}$ for $CuCl_2 \cdot 2H_2O$ and $10^{-6}$ for $MnF_2$. The fields of critical areas are 5.5 kOe and 91 kOe, correspondingly. It is seen, that the magnetostriction parameters and elastic constants are interrelated. The established bond of the energy of the elastic deforming stresses transformed by T,H,P with the processes of the structure changes and the state of electron bonds magnetic and resonance properties is realized both in the structural phase transitions and phase states.

The revealed regularities of IE stresses treated as the main cause of volume change under T-P-H effect through mechanisms of bulk elasticity as well as the relation of these mechanisms to changes in electric conduction let us to state that not all the electrons participate in conduction. Some of them are located at the filled shells bound to nodal points (ions) and forming the electron density and free electrons related to the processes of electric conduction. The mobility of electrons originates from the fact that the binding energy of electrons located from atom or molecular center to periphery is two and more orders of magnitude lower in the case of multi-component systems. Since the binding energy of electrons participating in the formation of lattice structure can be in a relative correspondence with the energy of IE stresses, and then namely the electrons can play a principal role in structure changes and be a defining factor of T-P-H thermodynamic influence.

As known, every single-crystalline or polycrystalline structure is formed by repetition of groups of ions, molecule, atoms in the bonds, thus forming sites, and the experiment shows that in lattice sites the atomic and molecular magnetism develops as magnetic non-compensation. Such non-compensation has its maximum at T=0 K and it goes to zero at high temperatures directly connected with processes of volume change and the reduction of the bond between sites and atoms.

Showing the role of the laws of cubic elasticity, we stress that if changes in volume are not accounted for completely, any methods of theoretical modelling remain an approximate solution with the empirical selection of coefficients to define changes in the exchange. The approximations taking these volume changes into account need much correlation.

2.7    For instance, the demonstrative is the result of intersection of approximated dependencies at the critical point $P_x'$ (Fig. 6a). The presence of three parameters (T-P-H) and the revealed regularities of their influence on the resistance has specified the approaches to substantiate the role of mechanisms of elastically deforming compression in the formation of structural phase transitions and the region of stress concentrators. The position of separated $P_x'$ critical point (Figs. 5, 6) shows the sequence of the physical process forming the region of PT in the form of a critical area fixing the jump of resistivity followed by the transition to conducting (metallic) phase. The changes of resistivity obey the law of linear elasticity. This result has made it possible to establish a very important factor of conformities in T-P-H ratio ($6.3^o$K-1 kbar-2.3 kOe) and to show that the backpressure and H decrease the level of IE stresses in the structure. The reduction of resistivity properties is analogous to the thermoelastic compression and heat sink (cooling).

Being the most prominent in regularities of hysteresis, differences of anisotropy, magnetoelasticity and thermoelasticity are revealed by linear changes of $T_{ms}(H)$ and $T_{ms}(P)$, Fig.6. Here we add a significant result of distinction of the temperature maxima of magneto-, baro-, magneto-baro-resistive effects $T_{PP} = T_{mc}$ (Fig. 6b) defining the reasons of SPT of type I and II. This is a straight confirmation of the formed structural phase transition. The physical processes are located in the region of thermoelasticity priorities. This follows from the revealed correspondences between T, H and P. The transition to the priorities to manetoelasicity and $T_c$ takes place with the increase of the field H.

One more important result adds to the role of bulk elasticity in linear magnetostriction dependences (Fig. 1a). It became possible to specify $P_x$, $PP_x$ critical points and the regular changes of the field-temperature dependence $T(H_g)$ (Fig. 1b) . That are the regularities of the dependence of complex processes of elastically deforming stresses in the structure that point to the separation of phase states limited by $P_x$ and $PP_x$. Here we also pay attention to the antiphase effect of temperature and to the formation of structural phase transition in the region where magnetoelastic stresses are of priority for H increasing; the mechanisms of thermoelasticity are involved with T decrease, and this is demonstrated by the revealed regularities. It becomes possible to approximate the magnetostriction properties on the dependencies for $T=0^\circ K$ and to show the role of mechanisms of magnetoelastic stresses under the maximal magnetic non-compensation resulting from the minimal volume for $T=0$ K. As a result, the position of PT has been determined and the role of magnetoelastic anisotropy has been shown to be defining $T_x$ ($H_x - 205$ kOe, $T=0$ K), Fig.1a,b. It follows that position of $T_c$ on this curve is, by definition, a fact of averaged conformity of ED stresses with the change of priorities, under the realization of structural phase transition between $T_{mc}(H=0, T_{PP})$ and $T_x(H_x, T=0)$.

Positions of critical points are connected with changes in volume, it is also a consequence of chemical composition, technology of preparation and the intrinsic magnetic non-compensation as well as the reaction of volume on T-P-H effect through the mechanisms of EAD stresses. It is worth to note the regularities in the hydrostatic pressure (P) effect (Fig. 1a), all the critical points are invariable for the composition under consideration, while the properties and position of PT dependence vary with respect to $T_x(H_x, T=0)$ (P and H influence are of single value).

The most demonstrative is the result where regularities in changes of properties and SPT of type I allow to register superposition of critical points $PP_x$ and $P_x$ (Figs. 2-5). The elastic regularities of T-P-H effect have been revealed. The location of these points shows that at the influence of two thermodynamic parameters, one of them should be varied in angular coordinates. This regularity gives more objective illustration of the physical processes studied in bulk samples.

The analysis of relationship between structural peculiarities and bulk elasticity in a real physical process has shown that the structural phase transition may exist at $T=0$ K. Moreover, the experiment gave the result close to that parameter. This effect is accompanied by a high jump of the properties, including heat capacity and susceptibility. Its parameters are $T_x$ ($H_x=6.5$ kOe for $T=0$ K), Fig. 2a,b. This means that the mechanisms of magnetoelastic ($H_x=6.5$ kOe) stresses correct the position of the structural phase transition for $T=0$ K. Next, the field-temperature dependence differentiates the phase states with the elements of superconducting and conducting phases. Here we note that $T_P$(H) dependence(Fig. 2a,b) of the PT of type I much differs from $T(H_g)$ (Fig. 1a,b) of PT of type II first of all by critical line position of different sign [10] and by existence of critical point P($H_P$, $T_P$) which has not been fixed earlier. This regularity of changes in phase transition is connected with position of point $PP_x$, in the results under consideration, and this is the defining factor for the formation of superconducting phase. It also shows the role of ED stresses in the effect of anomalous hysteresis within the region of structural phase transition. These results demonstrate the antiphase effect of the temperature on properties for separate phases.

The critical point of the phase transition at T=0 K for the given system functions as a triple point. In this region under thermo-ED compression, the mechanisms of elastic stresses start operating to realize a jump of properties when the region of phase transition is formed where the changes of the dynamics of the conduction-electron mobility are observed. Subsequently, structural phase transition results in considerable changes of heat capacity, resistivity, magnetization, thus characterizing the superconducting and conducting phases as they are.

Our attention is paid next to numerical parameters of conformal influence of T-P-H. We also note the position of the parameter where the region of field-temperature dependences $T_P$(H) and $T_P$(H,P) (Fig. 2a) is limited by P($H_P$, $T_P$) point with $T_P$=9.2 K. This critical point limits the region of existence of phase states, it can be determined for any structures by investigating properties and PT under the influence of high hydrostatic pressures.

The analysis of the role of $P_x'$ point where resistive dependencies intersect each other (Figs. 6a, 7a) in the physical process of volume change under T-P-H influence through mechanisms of ED compression is worth mentioning too. These results confirm regular involvement of the mechanisms of elasticity forming PT in the effect of colossal magnetoresistance being under study for many years.

With the novel approaches for analyzing the results of investigations, when the defining role of the effect of thermodynamic parameters T-P-H is considered in view of the mechanisms of EAD stresses, it is possible to explain numerous anomalies, peculiarities and effects by relationships between structure and elasticity.

### 3. Conclusions

Contemporary problems related to conducting magnet-containing media entail study of numerous experimental results obtained for systems of various compositions. Being based on the laws of bulk elasticity, new methods are to be used to analyze reversible physical processes induced by thermodynamic parameters.

The investigated objects are finite-size samples prepared by the conventional technology. With high-temperature annealing technique, the chemical compounds are formed with repeated sites and bonds in the structure. The atomic-electronic binding energy is optimally minimal in that process of high-temperature heat removal, while stresses are maximal. These are the structure-forming principles of energy distribution. Subsequent cooling directly relates to heat removal and the mechanisms of bulk thermo-elastic compression involved. An obvious evidence is the linearity of the lattice parameter within the temperature range of 300-1500 K [19] in $La_{0.875}Sr_{0.125}MnO_3$ and down to 0 K as shown by our analysis. This is a clear demonstration of the role of bulk elasticity in physical processes under the temperature, magnetic field or hydrostatic pressure influence. Such a consideration shows that any heat transfer is also a causal basis of the involvement of mechanisms which modify (expansion-compression) internal IE stresses. As a consequence, there follows the sign-alternating influence on structure change, i.e. heat application (heating) means the increase of IE stresses, heat removal (cooling) results in the decrease of ED stresses. From this standpoint, similar processes are observed when baroelasticity (P) and magnetoelasticity (H) mechanisms are involved as shown by the results of our analysis [20].

The relation between thermodynamic effect through heat-transfer processes transformed into mechanisms of elastically deforming stresses in structure changes has not been exposed finally, that is why their causal role in the formation of properties and PT has not been studied completely by model theoretical representations. This is because there in no complete understanding of relations between heat capacity (heat transfer) and IE stresses and because of uncertainty in substantiation of the role of these mechanisms and significance of the estimating priorities of energies in a variety of physical processes studied.

Being a complex inhomogeneous process with peculiar experimental linear regularities, elastic-anisotropic deforming stress can be taken into account with respect to power distribution by means of coefficients, but not always grounded. Moreover, the heat transfer, the heat conduction and IE stresses are bound in the structure of energies of mobile electrons, and linear changes observed in properties and PT determine the causal role of elasticity, more exactly of bulk elasticity in the processes of T-P-H effect.

When treating the heat transfer as a factor forming elastically deforming stresses in the structure and also redistributing the inner energy of interactions, we can conclude that in the laws of interactions and conformities one should take into account [17,20]:

a)    the energy of heat transfer transformed by mechanisms of elastically deforming stresses (partially accounted for in molecular-field models) resulting in structure changes;

b)    the energy of quantum-mechanical forces determined by the short-range atomic-electronic bonds (small contribution).

The T-P-H effect through the properties of elasticity varies the volume, redistributes the energies, thus forming the structural phase transition and properties in each phase state.

The analyzed processes of a fast (linear) increase of magnetization or magnetostriction in the magnetic field allowed us to suppose that particles (sites and bonds in the structure) are of great vector (magnetic moment) of magnetic non-compensation at definite conditions. This phenomenon results from the participation in structure formation of numerous individual non-compensated spins, bound electron density in intermolecular (modal) space. To our regret, it was not possible completely to explain the interactions or the role of elastically deforming stresses in non-compensative processes in theoretical molecular-field models.

Numerous results obtained for $LaMnO_3$, $CuCl_2 \cdot 2H_2O$, $Eu_{0.55}Sr_{0.45}MnO_3$, $La_{0.56}Ca_{0.24}Mn_{1.2}O_3$ are based on new representations, based on the existence of the direct correlation between structural changes and regularities of elastically deforming stresses under T-P-H effect. That fact defines the redistribution between interactions and conformities of the atomic-electronic binding energy and the energy of IE stresses in physical processes. In other words, this correlation of structure changes can be treated as the law determining the degree of conformity and correlation between heat flow (heat capacity) transformed by the energy of IE stresses. As noted, this correlation is always present.

The applied analytical methods allowed to emphasize the determining role of IE stresses in the formation of structural phase transitions of the first and the second type using the linear method of separation of locations of critical points and phase states at $0^o K$. This fact was a reason for finding generalizing principles named secondary indications which determine and fix the position of PT at 0 K. These indications are:

a)      unambiguous location of maxima of thermo-baro-resistive, thermo-baro-magneto-resistive effects at $T_{PP}=0$ K corresponding to the structural phase transition;

b)      single-phase influence of T and H on changes of magnetization properties at the initial section of the dependence and the antiphase influence on the same properties after the phase transition;

c)      revealed regularity of magnetization properties in the region of structural phase transition into inverse (anomalous) hysteresis effect;

d)      a sign-alternating effect of temperature on magnetization properties in different phases ;

e)      succession in positions of mentioned points $PP_x$ and $P_x$ and their direct relation to the field-temperature dependence $T_P(H)$, the boundary separating the superconducting and conducting phase states.

For a more complete understanding of the laws of bulk elasticity in physical processes it is very important to analyze the significance of the mentioned critical lines and points for the formation of structural phase transitions and properties.

The presented generalizing analysis shows the correspondence and interaction of binding energy and elastic energy. The neglection of this energy in the course of model theoretical analysis results in practical refusal of an important part of the energy which forms and changes the phase transition and properties.

**Acknowledgements.** The author expresses sincere thanks to N.P. Boyko and E.V. Sapego for the interest, attention and help in the work and T.N. Melnik for attention and valuable suggestions.

## Figure captions

**Fig.1**

a) Field dependences of the longitudinal magnetostriction for $LaMnO_3$;

b) Field-temperature dependence $T_p(H_g)$ of PT.

c) Dependence of magnetization change in hysteresis effect.

**Fig. 2**

a) Field dependences of longitudinal magnetostriction for $CuCl_2 \cdot 2H_2O$;

b) Field-temperature dependence of changes in the phase transition $T_P(H)$, $T_p$ $(H, P)$ for P = 11,2 kbar in $CuCl_2 \cdot 2H_2O$;

c) Dependence of magnetization change in the effect of inverse hysteresis in $CuCl_2 \cdot 2H_2O$.

**Fig. 3**

Field-temperature dependence of changes in $T_P(H)$ - the phase transition in $Sm_{1-x}Sr_xMnO_3$.

**Fig. 4**

Field-temperature dependence of changes in phase transition $T_p(H_g)$ in $Eu_{0.55}Sr_{0.45}MnO_3$.

**Fig. 5**

a) Dependence of magnetoresistive effects MR (%);

b) Field-temperature dependence of changes in phase transition $T_p(H_g)$.

**Fig. 6**

a) Temperature dependence of resistivity of $La_{0.56}Ca_{0.24}Mn_{1.2}O_3$ ceramic sample: 1 - P = 0 kbar; 2 - P = 0, H = 8 kOe; 3 - P = 6 kbar; 4 - P = 12 kbar; 5 - P = 18 kbar; 6 - P = 18 kbar, H = 8 kOe;

b) Temperature dependence of baroresistive, baromagnetoresisitive, magnetoresistive effects: 1 - H = 8 kOe, P - 18 kbar; 2-P-O, H = 8 kOe; 3 - P = 12 kbar; 4 - P = 6 kbar; 5 - P = 6 kbar H = 8 kOe; 6 - H = 8 kOe.

**Fig. 7**

a) Temperature-field dependence of phase transition with critical point $P_x$;

b) The charge-ordered phase of various compounds $Re_{1/2}Ae_{1/2}MnO_3$ plotted on the magnetic field-temperature plane. [8]

**Fig.8**

Temperature-field dependence of phase transition with critical point $PP_x$ of Nd1-xCa$_x$MnO$_3$ (x=0.45) [15].

**Fig. 9**

a) The stability of linear changes in $La_{0.7}Sr_{0.3}MnO_3$ in whole temperature range 350 – 1500 K.

b) Monotonous linear increase of crystal lattice parameter of $LaMnO_3$ and $La_{0.7}Sr_{0.3}MnO_3$ polycrystalline samples [26].

**Fig. 10**

Dependence of interplanar spacings $d_{400}$, $d_{040}$, $d_{003}$ on uniform compression in $CuCl_2 \cdot 2H_2O$ [4].

**Fig. 11**

The magnetoresistivity in copper at fixed temperatures: $T_1$=86 K, $T_2$=63 K, $T_3$=20 K.

**Fig. 12**

a, a') Magnetic phase diagram $MnF_2$ in the field parallel to (001) axis.

b) Temperature and field dependence of a structural phase transition of the first order with separated critical point $PP_x$ and $T_{pp}$ ($T_x$=0, $H_x$~93 kOe).

c, c') The jump of magnetization of $MnF_2$ at varied magnetic field orientations with respect to (001) axis at T=4.2 K.

d) Frequency and field dependence of resonance absorption of $MnF_2$ in the field parallel to (001) axis at T=4.2 K.

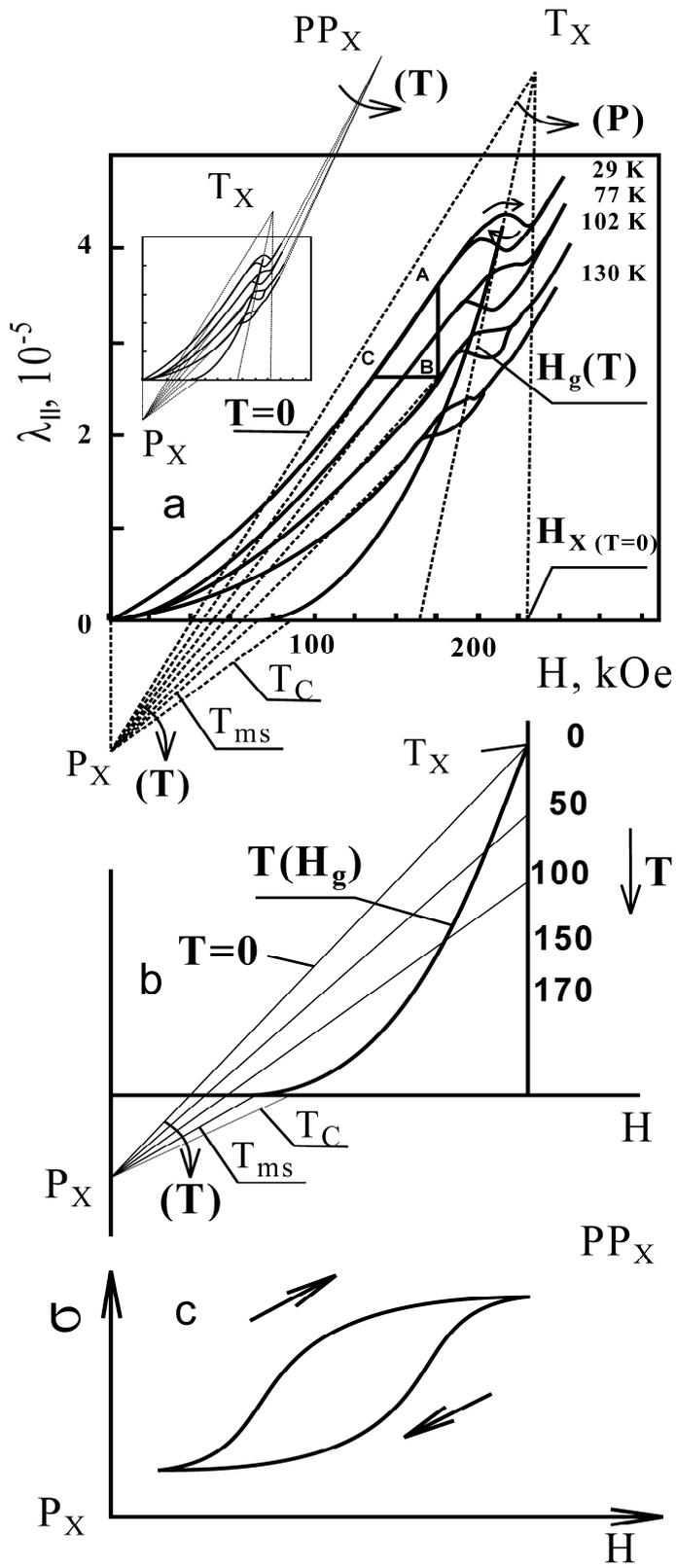

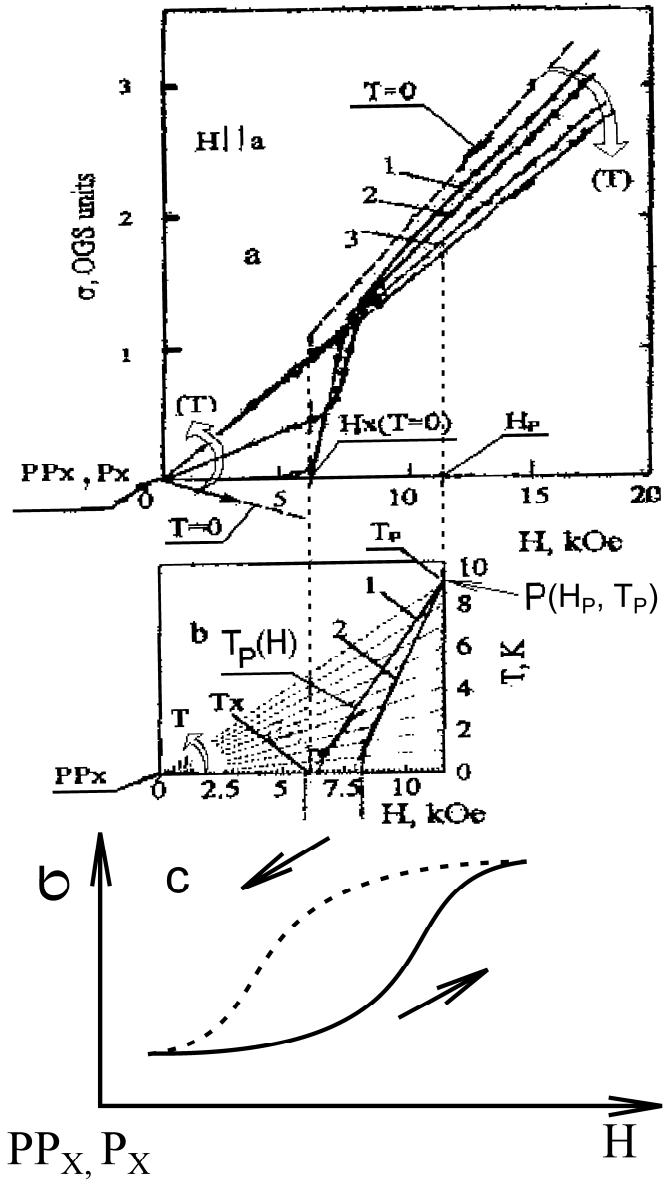

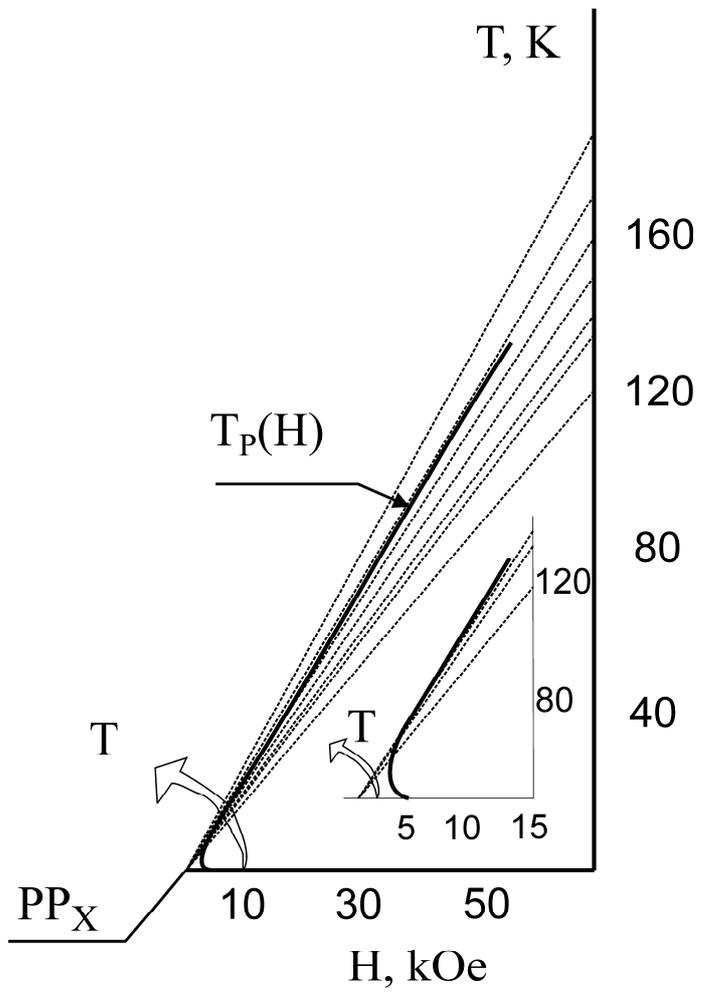

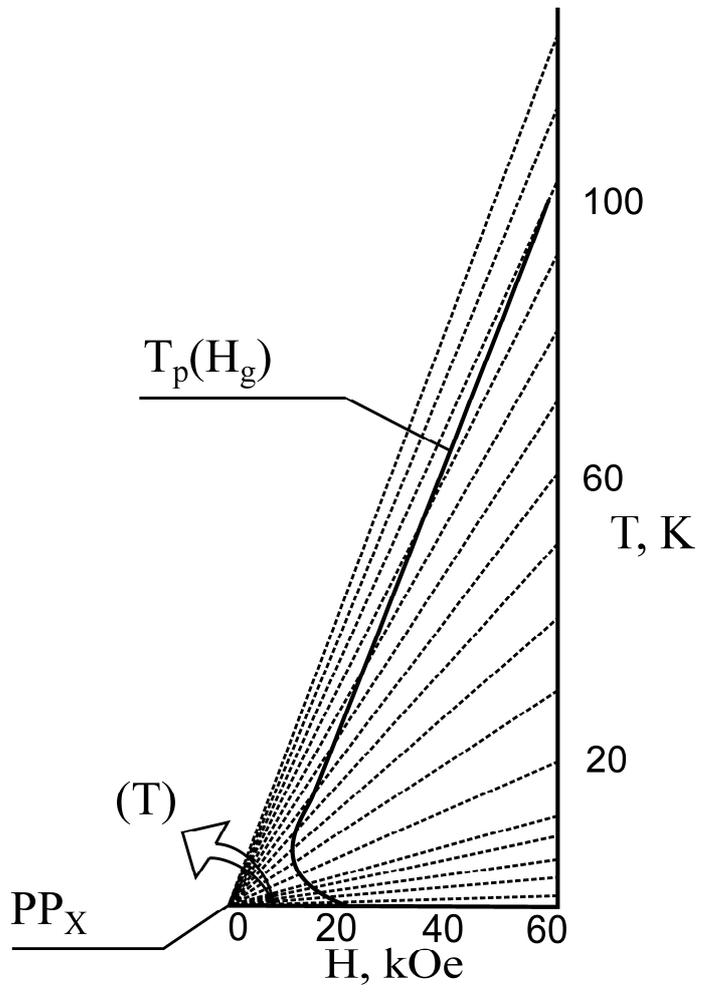

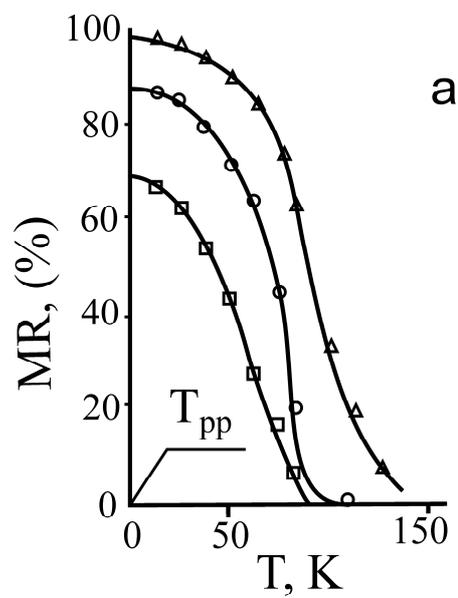

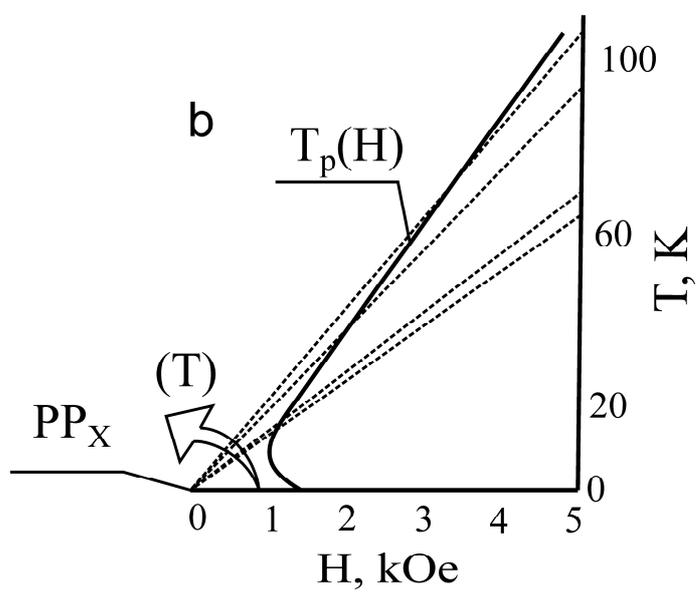

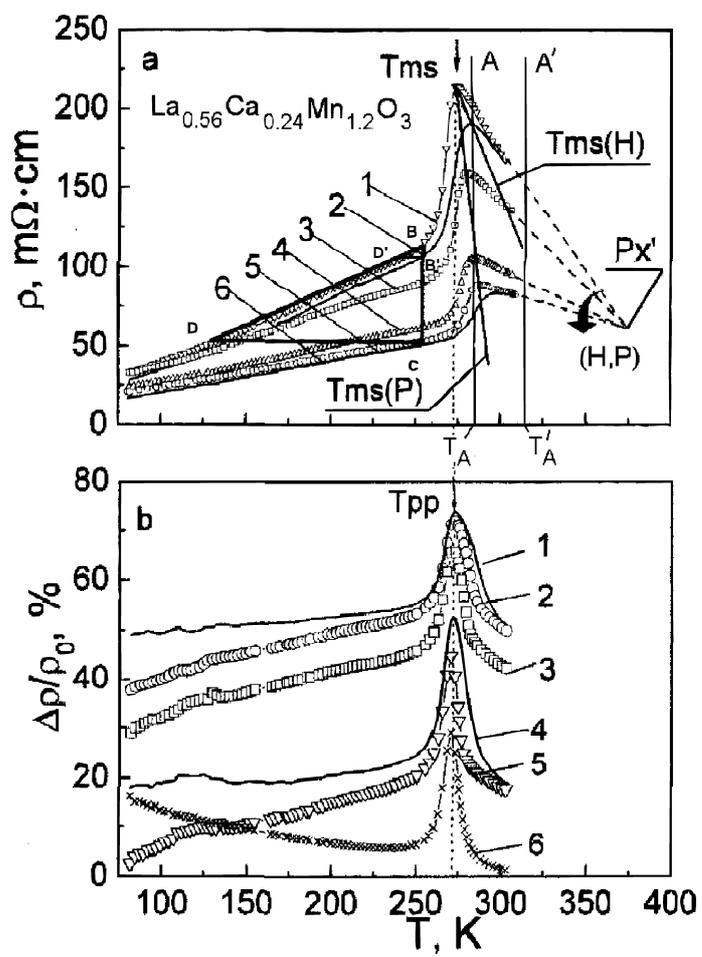

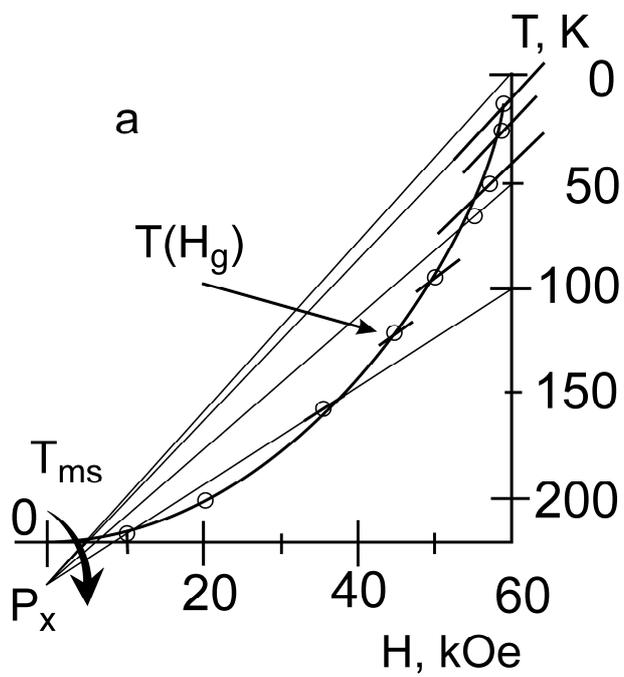

a

T, K

T(H_g)

T_ms

0

P_x

H, kOe

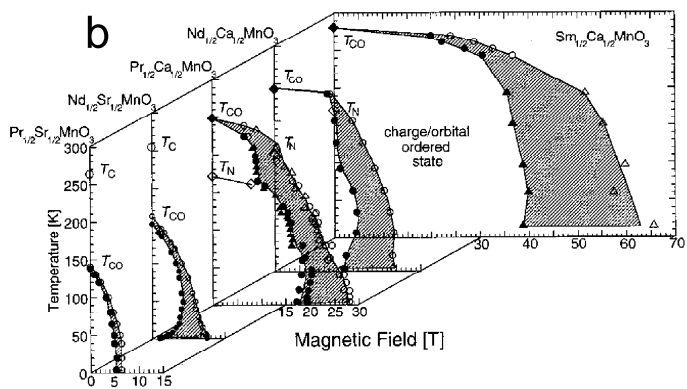

b

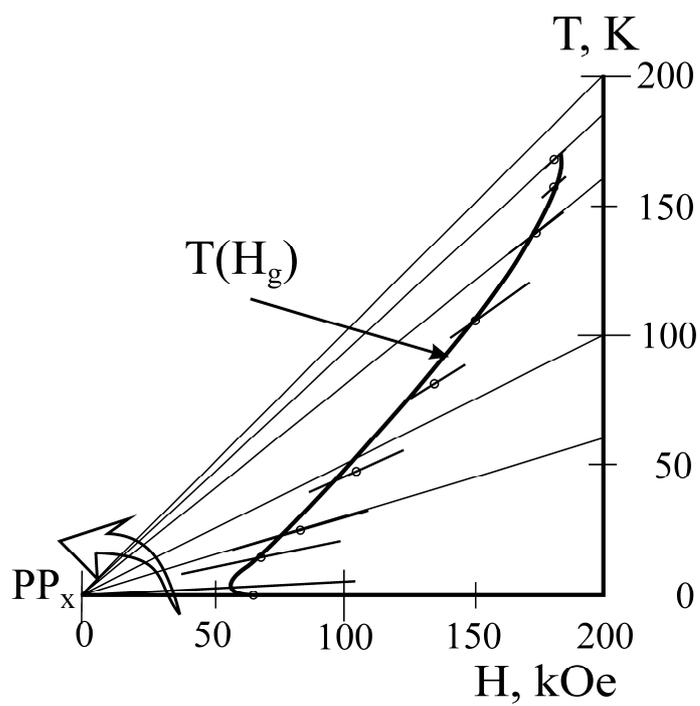

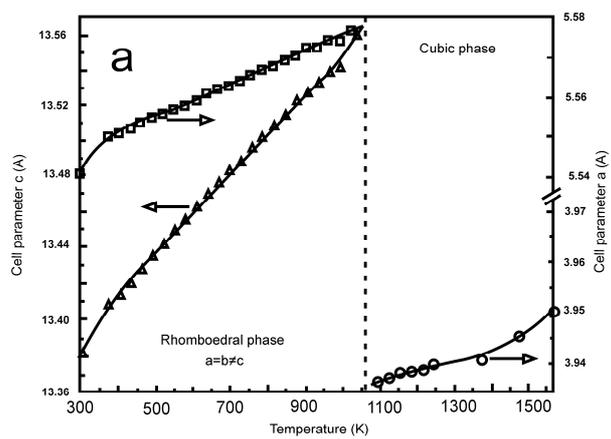

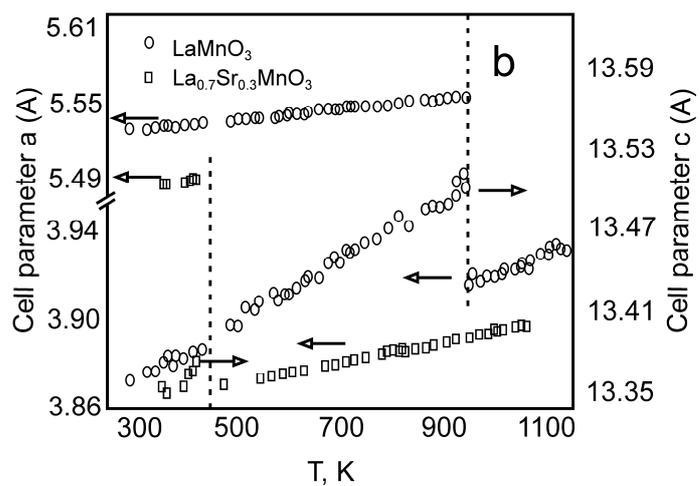

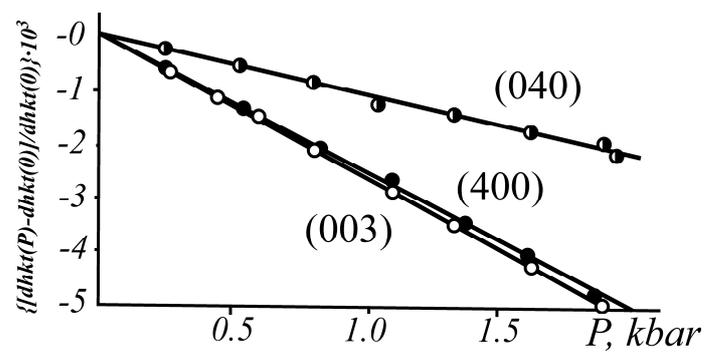

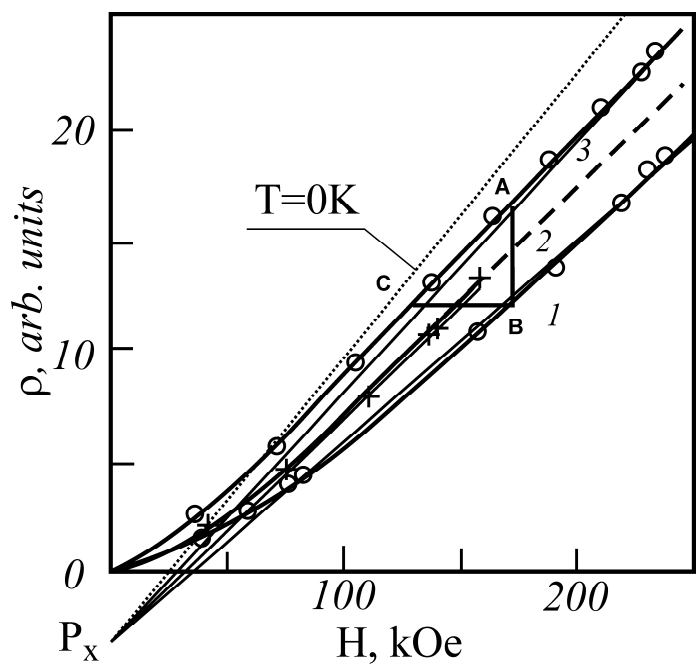

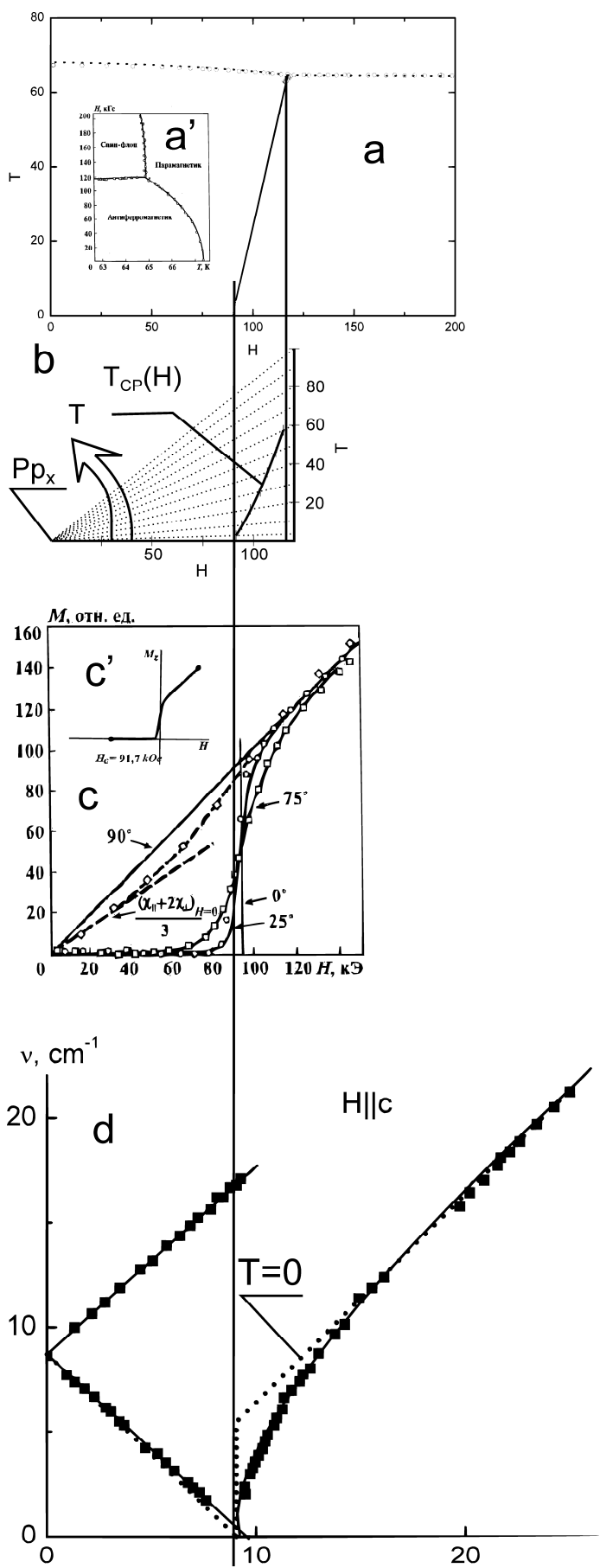